\documentclass[aps,prl,twocolumn]{revtex4-1}

\usepackage[TS1,OT1,T1]{fontenc}
\usepackage{graphicx}
\usepackage{amssymb}

\begin{document}
\title{Observation of Bose-Einstein Condensation in a Strong Synthetic Magnetic Field}
\author{Colin J. Kennedy$^{*}$}
\affiliation{MIT-Harvard Center for Ultracold Atoms, Research Laboratory of Electronics, Department of Physics, Massachusetts Institute of Technology}
\author{William Cody Burton}
\affiliation{MIT-Harvard Center for Ultracold Atoms, Research Laboratory of Electronics, Department of Physics, Massachusetts Institute of Technology}
\author{Woo Chang Chung}
\affiliation{MIT-Harvard Center for Ultracold Atoms, Research Laboratory of Electronics, Department of Physics, Massachusetts Institute of Technology}
\author{Wolfgang Ketterle}
\affiliation{MIT-Harvard Center for Ultracold Atoms, Research Laboratory of Electronics, Department of Physics, Massachusetts Institute of Technology}
\date{\today}
\begin{abstract}
Extensions of Berry's phase and the quantum Hall effect have led to the discovery of new states of matter with topological properties. Traditionally, this has been achieved using gauge fields created by magnetic fields or spin orbit interactions which couple only to charged particles. For neutral ultracold atoms, synthetic magnetic fields have been created which are strong enough to realize the Harper-Hofstadter model. Despite many proposals and major experimental efforts, so far it has not been possible to prepare the ground state of this system.  Here we report the observation of Bose-Einstein condensation for the Harper-Hofstadter Hamiltonian with one-half flux quantum per lattice unit cell.  The diffraction pattern of the superfluid state directly shows the momentum distribution on the wavefuction, which is gauge-dependent. It reveals both the reduced symmetry of the vector potential and the twofold degeneracy of the ground state. We explore an adiabatic many-body state preparation protocol via the Mott insulating phase and observe the superfluid ground state in a three-dimensional lattice with strong interactions.
\end{abstract}

\maketitle

Topological states of matter are an active new frontier in physics. Topological properties at the single particle level are well understood; however, there are many open questions when strong interactions and correlations are introduced \cite{WenSCI2012,SenthilSCI2014a} as in the $\nu = 5/2$ state of the fractional quantum Hall effect \cite{EnglishPRL1987} and in Majorana fermions \cite{KanePRL2008,XuNANO2012,YazdaniSCI2014}.  For neutral ultracold atoms, new methods have been developed to create synthetic gauge fields.  Forces analogous to the Lorentz force on electons are engineered either through the Coriolis force in rotating systems \cite{DalibardPRL2000a, KetterleSCI2001a, GemelkeARV2010}, by phase imprinting via photon recoil \cite{ZollerNJP2003a, Kolovsky2011bh, SpielmanNAT2009a, BlochPRL2013a, KetterlePRL2013a}, or lattice shaking \cite{StruckNPH2013a, EsslingerNAT2014a}.  Much of the research with ultracold atoms has focused on the paradigmatic Harper-Hofstadter (HH) Hamiltonian, which describes particles in a crystal lattice subject to a strong homogeneous magnetic field \cite{Harper1955, AzbelSPJ1964a, Hofstadter1976a}. For magnetic fluxes on the order of one flux quantum per lattice unit cell, the radius of the smallest possible cyclotron orbit and the lattice constant are comparable, and their competition gives rise to the celebrated fractal spectrum of Hofstadfer's butterfly whose sub-bands have non-zero Chern numbers \cite{BlochNAT2015} and Dirac points.

So far, it has not been possible to observe the ground state of the HH Hamiltonian, which for bosonic atoms is a superfluid Bose-Einstein condensate.  It is unknown whether this is due to heating associated with technical noise, non-adiabatic state preparation, or inelastic collisions. These issues are complicated, since all schemes for realizing the HH Hamiltonian use some form of temporal lattice modulation and therefore are described by a time-dependent Floquet formalism. The HH model arises after time averaging the Floquet Hamiltonian, but it is an open question to what extent finite interactions and micromotion lead to transitions between Floquet modes and therefore heating \cite{GoldmanAXV2014a, BukovPRA2014a, CooperPRA2014, MuellerARV2014}. Bose-Einstein condensation has been achieved in staggered flux configurations \cite{BlochPRL2011a, StruckNPH2013a} and in small ladder systems \cite{AtalaNPH2014a, SpielmanAXV2015a} further highlighting its noted absence in the uniform field configuration.

In this article, we report the first observation of Bose-Einstein condensation in Hofstadter's  butterfly. The presence of a superfluid state in the HH lattice allows us to analyze the symmetry of the periodic wavefunction by self-diffraction of coherent matter waves during ballistic expansion -- analogous to a von Laue x-ray diffraction pattern, which reveals the symmetry of a lattice. Using this method, we show that the unit cell of the magnetic lattice is larger than that of the underlying cubic lattice. This reflects that the vector potential necessarily has a lower symmetry than a uniform magnetic field, since a transitionally invariant field can only be realized by a vector potential that breaks this translational invariance. In our experiment, we analyze the wavefunction directly and observe both its non-gauge invariance \cite{DasSarmaPRA2011} and ground state degeneracy. Finally, we explore a many-body adiabatic state preparation protocol where the many-body gap of the Mott insulator is used to switch superfluid order parameters without creating excess entropy. We realize the HH Hamiltonian superfluid ground state in a three-dimensional lattice with strong interactions.

\begin{figure*}
\centering
\includegraphics[width=\textwidth]{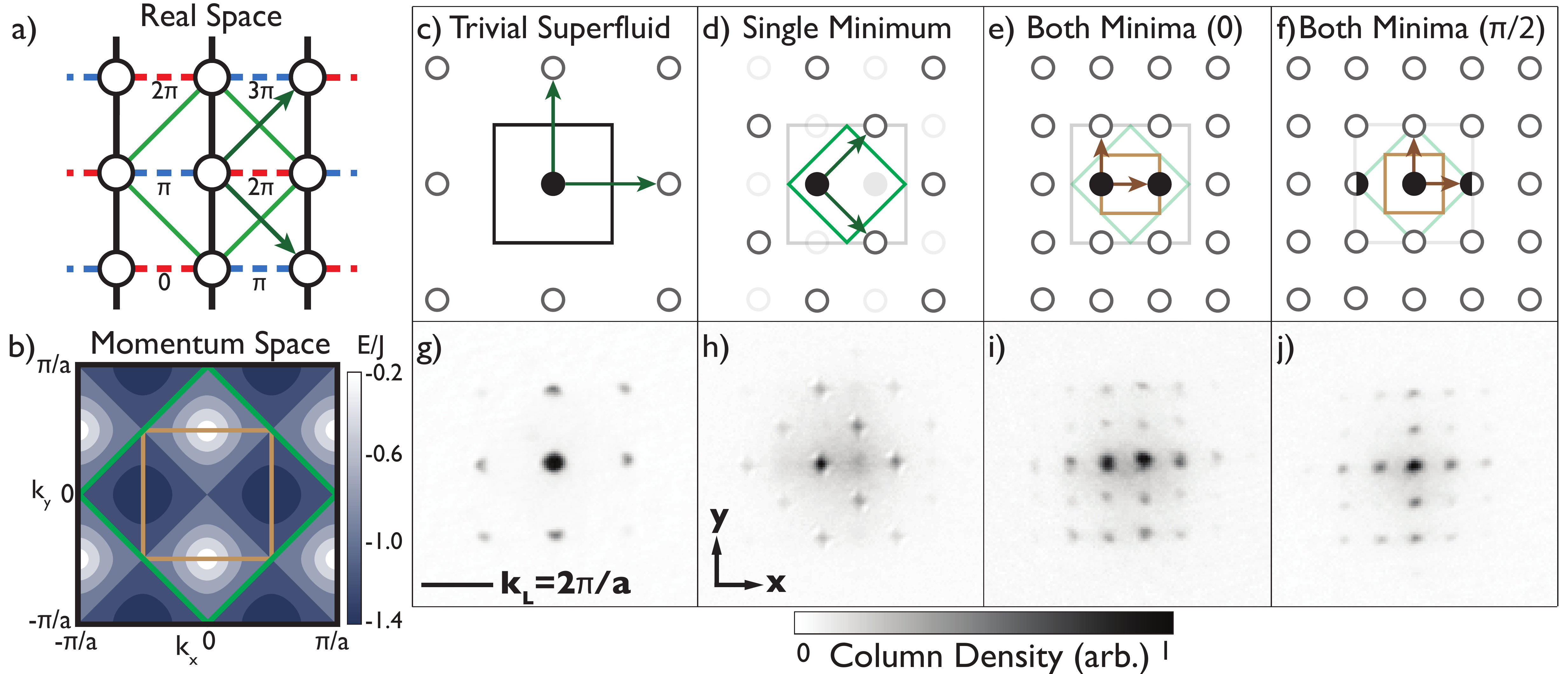}
\caption{Observation of Bose-Einstein condensation in the Harper-Hofstadter model with one half flux quantum per unit cell.  (a) Spatial structure of the cubic lattice with the synthetic vector potential -- (dashed) x-bonds feature a spatially dependent tunneling phase $\phi_{mn} = \pi(m+n)$, whereas tunneling along (solid) y-links is the normal tunneling. The synthetic magnetic field generates a lattice unit cell that is twice as large as the bare cubic lattice (green diamond). (b) The band structure of the lowest band of the HH Hamiltonian shows a twofold degeneracy of the ground state and two Dirac points located at $k_y=\pm\pi/2a$. The magnetic Brillouin zone (green diamond) has half the area of the original Brillouin zone. Due to degeneracies related to the indistinguishability of the sites in the magnetic unit cell, the primitive cell of the band structure is even smaller (doubly reduced Brillouin zone, brown square). These lattice symmetries are both revealed in time-of-flight pictures showing the momentum distribution of the wavefunction.
(c-f) Schematics of the momentum peaks of a superfluid. The dominant momentum peak (filled circle) is equal to the quasimomentum of the ground state. Due to the spatial periodicity of the wavefunction, additional momentum peaks (empty circles) appear, separated by reciprocal lattice vectors (green arrows) or vectors connecting degenerate states in the band structure (brown arrows). 
(g-j) Time-of-flight images. The superfluid ground state of the normal cubic lattice is shown in (g) in comparison to the superfluid ground state of the HH lattice (h-j).  In (h), only one minimum of the band structure is filled, directly demonstrating the fundamental symmetry of the HH Hamiltonian in our chosen gauge. The number of momentum components in (i-j) is doubled again due to population of both degenerate ground states of the HH Hamiltonian. The micromotion of the Floquet Hamiltonian is illustrated in (e-f, i-j) as a periodically shifted pattern in the x-direction, analogous to a Bloch oscillation. All diffraction images were taken after at least 30 ms hold in the HH lattice. The field of view of all experimental pictures is $631\ \mathrm{\mu m}\times631\ \mathrm{\mu m}$.}
\end{figure*}

The HH model with large synthetic magnetic fields is realized as an effective Hamiltonian engineered by laser-assisted tunneling processes in a tilted lattice potential. As in Refs. \cite{BlochPRL2013a,KetterlePRL2013a}, tunneling in the x-direction of a two-dimensional optical lattice is suppressed by an energy offset and then subsequently restored with a resonant Raman process with a momentum transfer, $\delta\vec{k} = k_x \hat{x} + k_y \hat{y}$. The z-direction consists of loosely confined tubes of condensate, so interactions are weak. In the time-averaged picture, this gives rise to the Hamiltonian:
\begin{equation}
\label{Harper}
H = \sum_{m,n} \left( -K\:e^{-i\phi_{m,n}}\hat a^{\dagger}_{m+1,n}\hat a_{m,n} -J\:\hat a^{\dagger}_{m,n+1}\hat a_{m,n} \right) + \text{H.c.}
\end{equation}
where $\phi_{m,n} = m k_x a + n k_y a = \pi(m+n)$ reflects the specific gauge implemented (Fig. 1a). An atom that travels around a single lattice unit cell picks up a phase of $\pi = k_y a$, with $a$ being the lattice spacing, so the Hamiltonian is equivalent to that of a charged particle in a magnetic field with one half of a magnetic flux quantum per unit cell. 

After preparing the atoms in the HH lattice (see Methods), we observe the momentum distribution of the wavefunction by suddenly turning off all laser beams to allow 20 ms time-of-flight, and measure the resulting density distribution with absorption imaging (Fig. 1h-j). We first note that the images show sharp peaks, which are the hallmark of a superfluid Bose-Einstein condensate in a periodic potential. This demonstrates for the first time that we have successfully prepared a low entropy state in the bulk HH Hamiltonian. In addition, the diffraction images directly show a reduced symmetry of the superfluid, despite the translational symmetry of both the lattice and the homogeneous synthetic magnetic field. 

Fundamentally, the vector potential is responsible for the broken translation symmetry of the lattice, and the time-of-flight patterns can be understood by examining these symmetries. In real space, there are two relevant unit cells: one is the unit cell of the cubic lattice, and the other is the unit cell of the Hamiltonian determined by the experimental gauge, which we call the magnetic unit cell. In highly symmetric gauges -- such as our experimental gauge and the Landau gauge -- a magnetic flux of $\alpha = p/q$ has a magnetic unit cell that is $q$ times larger than the original unit cell and contains $q$ indistinguishable sites \cite{DasSarmaPRA2011}. In momentum space, these unit cells correspond to the Brillouin zone of the underlying lattice and to the magnetic Brillouin zone, respectively.

Due to the indistinguishability of the sites in the magnetic unit cell, every quasimomentum state within the magnetic Brillouin zone is $q$-fold degenerate with other quasimomentum states connected by the translation symmetry operators of the original lattice modified by a phase $2\pi/q$. This modified translation symmetry connects all degenerate ground states in the magnetic Brillouin zone, and corresponds to a third relevant scale $2\pi/(qa)$ useful for understanding the diffraction pattern of a superfluid in the HH Hamiltonian. This length scale defines the doubly reduced Brillouin zone, which is the primitive cell of the band structure \cite{DasSarmaPRA2011}.

Experimentally, we observe the gauge-dependent diffraction pattern directly when the condensate occupies only one minimum of the band structure (Fig. 1h). The pattern's symmetry reflects that of the gauge potential (Fig. 1d), directly demonstrating the symmetry of the reciprocal lattice vectors $a(\hat{x}\pm\hat{y})$ of the magnetic Brillouin zone. A common belief is that all observables are gauge-independent. However, gauge-dependent observations can be made in time-of-flight images of ultracold atoms when the momentum distribution of the wavefunction is observed. The sudden switch-off of all laser beams transforms canonical momentum, which is gauge dependent, into mechanical momentum, which is readily observed \cite{SpielmanNPH2011a}.

The diffraction pattern shown in Fig. 1i reveals the symmetry of the band structure (Fig. 1b). In the experimental image, both degenerate states are equally populated and the peaks can be regarded as the sum of the diffraction pattern in Fig. 1h with an additional copy separated by the gauge-invariant scale $2\pi/(qa) = \pi/a$ in either $\hat{x}$ or $\hat{y}$ directions. The diffraction pattern reflects this two-fold multiplicity of momentum peaks in both the x and y directions and features a four-fold increase of visible peaks \cite{ZaleskiPRA2013}. To the best of our knowledge, this work is the first direct observation of the different symmetries characterizing a simple lattice with a homogeneous magnetic field: Figs. 1g-i show the symmetry of the lattice, the gauge potential, and the band structure. Realization of a Bose condensed ground state allows direct observation of both gauge-dependent and gauge-independent symmetry breaking.

A final feature of the observed diffraction pattern is the periodic oscillation of the peaks along the x-axis. This is the result of micromotion coming from acceleration due to the tilt and Raman drive. In the limit of a low drive amplitude, this micromotion is a Bloch oscillation in the tilted lattice. More generally, it is the effect of a time-dependent transformation between the rotating frame in which the HH Hamiltonian is realized and the lab frame in which it is observed \cite{KetterlePRL2013a}. Figs. 1e,i and f,j show diffraction patterns for the micromotion at zero phase and at $\pi/2$ phase -- equivalent to zero and $\pi$ phases in a Bloch oscillation due to the doubling of the unit cell. In the experiment, the absolute phase of the micromotion is a random variable due to wavelength scale drifts of the relative positions of the stationary and Raman lattices.

\begin{figure}
\centering
\includegraphics[width=\columnwidth]{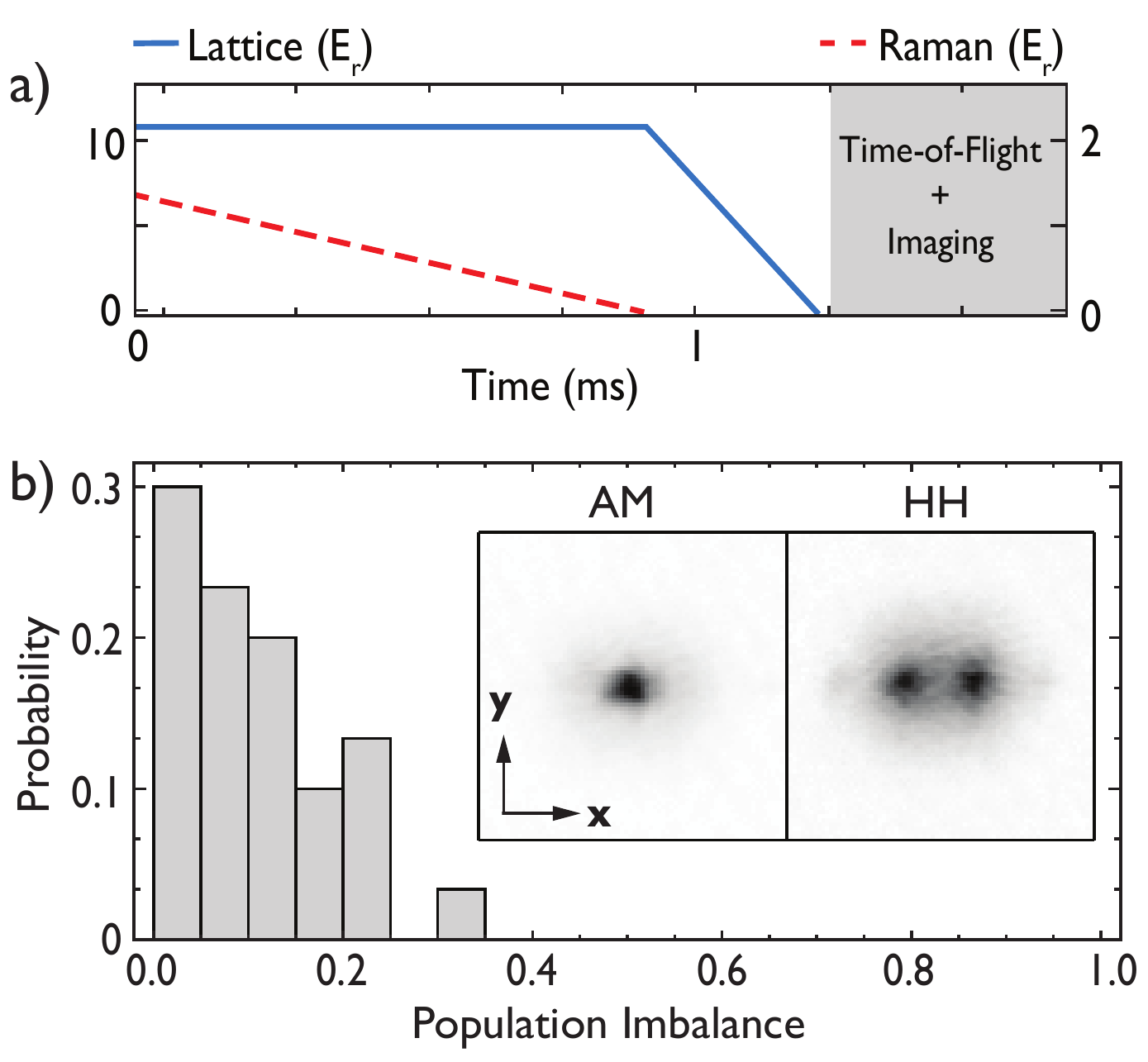}
\caption{Population imbalance of the two ground states of the HH Hamiltonian with 1/2 flux. (a) Band mapping sequence adiabatically connecting quasimomentum to free space momentum. The Raman beams were ramped down in 0.88 ms, followed by a linear ramp down of the lattice beams in 0.43 ms. (b) The histogram shows the relative population imbalance of the two degenerate minima, consistent with a spontaneous symmetry breaking mechanism during state preparation. The data consists of 30 shots taken after a hold time of 29.4 ms in the HH lattice. The inset displays a raw image for the band mapping of the 1/2 flux superfluid with two degenerate ground states compared to a topologically trivial superfluid with one ground state.}
\end{figure}

Qualitatively, almost all images show roughly equal population of both degenerate ground states. Occupation of only one state (as in Fig. 1h) is observed in less than 1\% of the shots. For a quantitative determination of the populations, we developed a band mapping technique that adiabatically connects the HH ground states to momentum states (see Supplement). Band mapping reduces the complicated diffraction patterns of Fig. 1 to two peaks, one for each degenerate ground state. Fig. 2b shows that the minima are most frequently populated with equal proportion, demonstrating the degeneracy of the minima and the robustness of the loading procedure to technical fluctuations. However, mean-field calculations predict that a superposition state, which would create spatial density modulation, is energetically unfavorable due to repulsive interactions \cite{DasSarmaPRA2011}. Therefore, the simultaneous occupation of both minima most likely indicates the presence of domains, composed of atoms in only one of the dispersion minima.

To rule out non-adiabatic transitions to higher bands as an alternative explanation of the two peaks in band mapping, we tested the technique on a topologically trivial Floquet superfluid (see Methods). This system consists of a cubic lattice with a tilted potential where tunneling in the tilted axis is restored by amplitude modulation (AM) \cite{NagerlPRL2010a, AlbertiNJP2010}. Its micromotion, which is similar to that of the HH system, and its trivial band structure allows identification of non-adiabatic excitations attributed to the micromotion. The resulting momentum distributions for both the AM and the HH systems are shown in Figure 2b. Non-adiabatic band excitations would appear as additional peaks in the tilted direction, which are not observed. Therefore, we interpret the band mapping procedure in the HH system as an adiabatic process.

\begin{figure}	
\centering	
\includegraphics[width=\columnwidth]{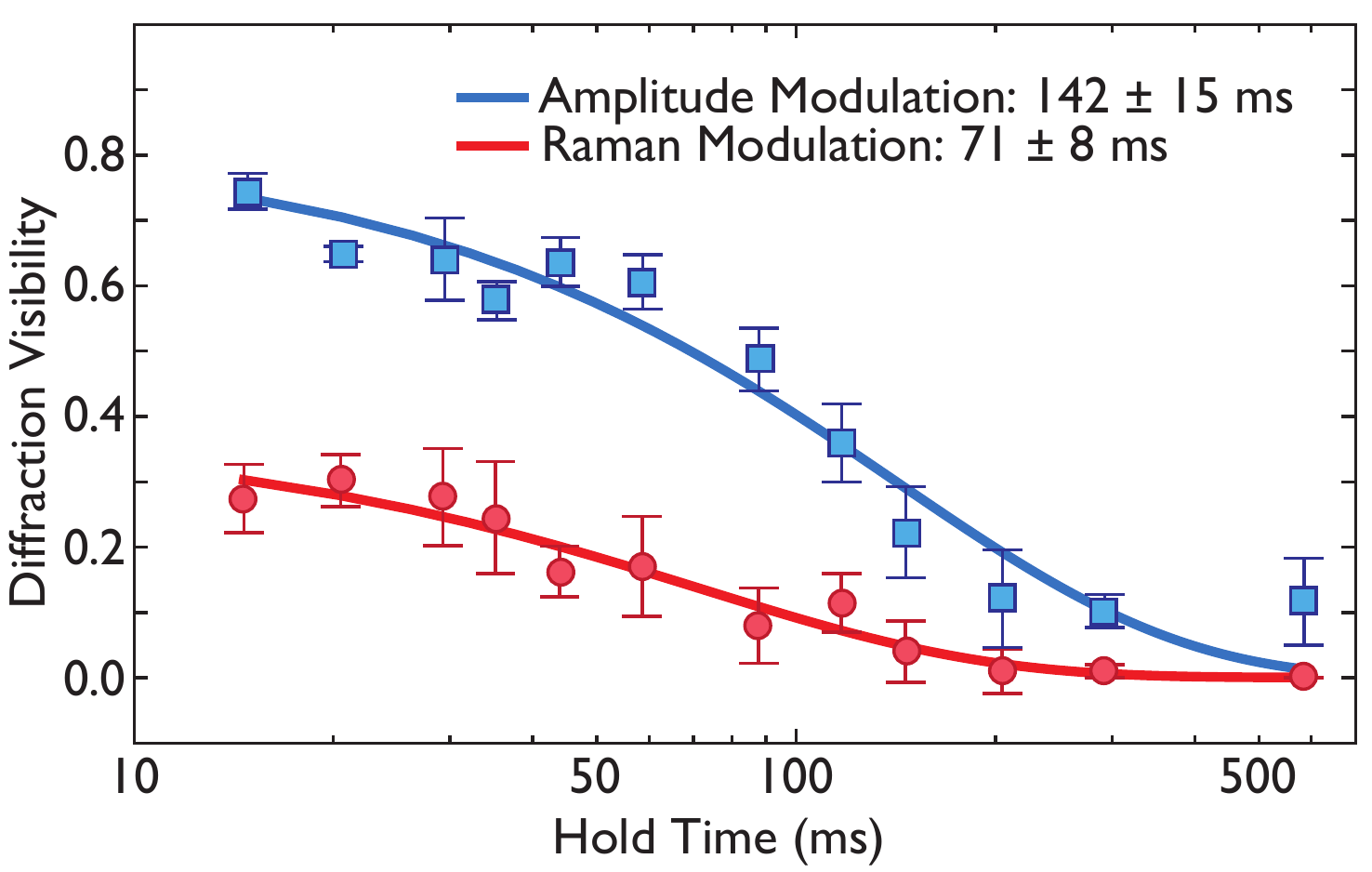}
\caption{Decay of Bose-Einstein condensates in modulated lattices.  The figure compares the decay of the 1/2 flux HH superfluid (red circles) to the decay of the AM superfluid (blue squares). Note that the lower visibility of the HH superfluid is due to the peak doubling, which at the same condensate fraction, leads to lower visibility. Exponential fits to the decay of the visibility of the diffraction patterns give lifetimes of 142 $\pm$ 15 ms and 71 $\pm$ 8 ms, respectively. The data were taken after a 10 ms hold time after switching on the final Hamiltonian using the non-adiabatic procedure (see Methods). Uncertainty is given by the statistical error of the mean of five repetitions of the experiment, added in quadrature to uncertainty in the peak visibility fitting.}
\end{figure}

The above results confirm many aspects of the weakly interacting superfluid ground state of the HH Hamiltonian. However, there are many open questions about topological physics in the presence of strong interactions, often involving small energy differences. To study this regime, long coherence lifetimes are necessary.

In order to evaluate our coherence lifetime and disentangle different sources of decoherence and heating, we extensively used the AM superfluid as a benchmark. Initially, the HH superfluid had a vastly shorter lifetime than the AM superfluid. Subsequent improvements in the stability of magnetic fields, beam pointing, Raman phase, and gradient alignment to the lattice direction improved the HH lifetime until both systems had a comparable lifetime in the ground state, given in Fig. 3. The lifetimes for the number of trapped atoms is much longer than the coherence lifetime in either case, with no discernible loss of atom number up to 500 ms. This leads to the conclusion that two-body dipolar interactions and three-body recombination, which would lead to particle loss from the trap, are not limiting the lifetime of the ground state.  The same applies to excitations to higher lying bands which, due to the tilt, are strongly coupled to the continuum via Landau-Zener tunneling. Therefore the decay in Fig. 3 is dominated by transitions within the lowest band, which can be caused by technical noise or by elastic collisions that transfer energy from the micromotion into heat. While our heating seems to be largely technical, it is useful to briefly discuss these collisions as an ultimate limit on the lifetime.

For the tilted lattice, one decay path is via overlap of neighboring Wannier-Stark states with offset energy $\Delta$, which can be transferred to excitations of the lowest bands, or to the free particle motion along the tubes orthogonal to the 2D lattice.  Such processes are described by Fermi's golden rule for transitions between Floquet states \cite{CooperPRA2014, MuellerARV2014}.  Using this framework, we derive the scattering rate for the AM superfluid to be between $\Gamma \sim 0.30-0.68\ \mathrm{s}^{-1}$, where the uncertainty comes from the uncertainty in our density measurement due to redistribution during lattice ramp-up. This estimate is smaller than the observed decay rate by a factor of $\sim10-20$. See Supplement for a derivation and experimental parameters.

In principle, an adiabatic procedure to prepare the ground state of the HH Hamiltonian should enable better control and higher fidelity in the final state. However, we empirically found that a sudden turn-on of the tilt and the Raman beams gave the most consistent high contrast images, and was therefore useful in evaluating technical improvements. A slower turn-on could introduce more technical heating, and the entropy created in the sudden turn-on could be absorbed by other degrees of freedom. Such an entropy reservoir is provided in the 2D lattice schemes by the third dimension -- in our case tubes of condensates with $\sim$500 atoms. For extensions to 3D lattices, we implement an adiabatic state preparation procedure.

\begin{figure}	
\centering	
\includegraphics[width=\columnwidth]{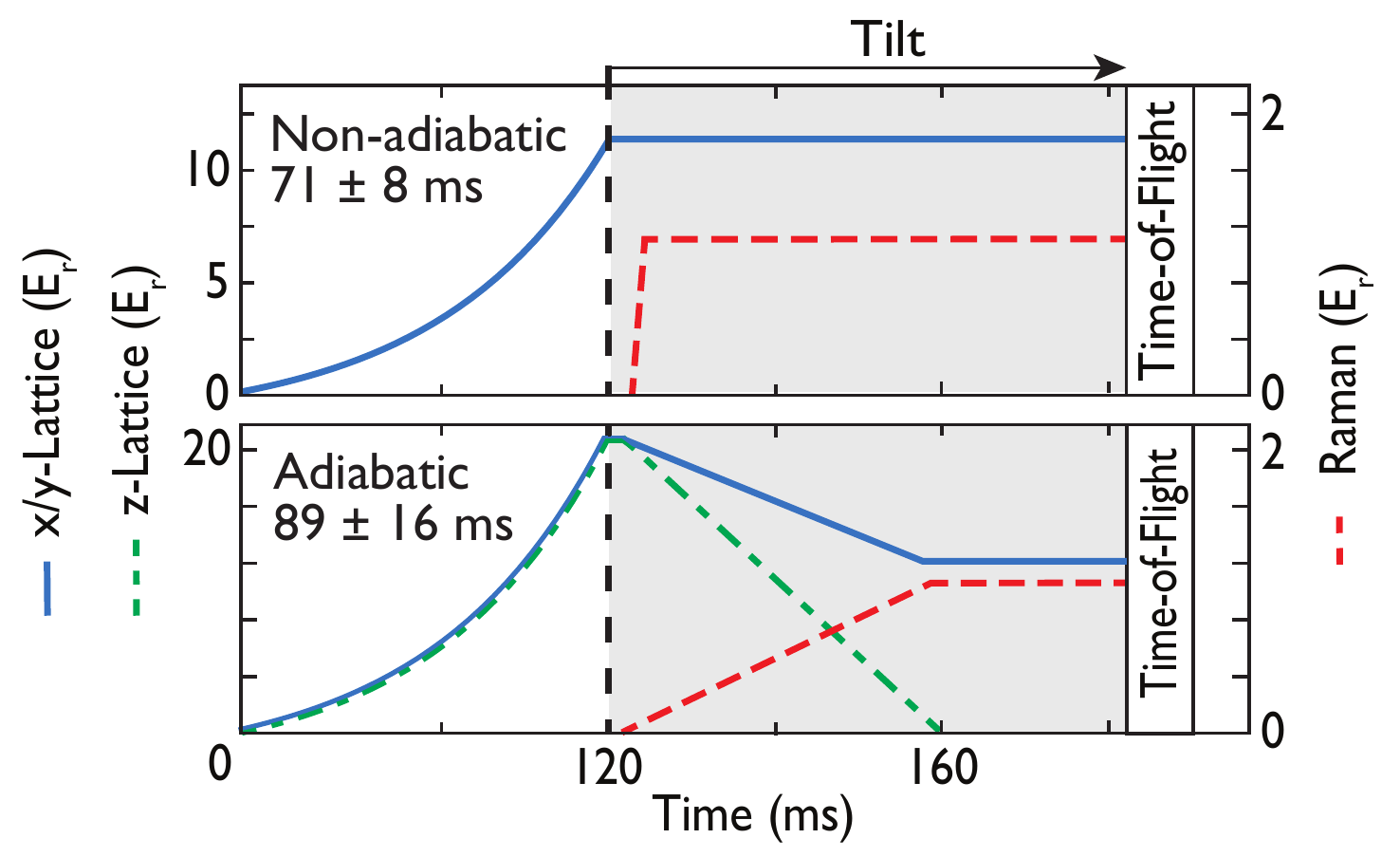}
\caption{Experimental sequences for two different state preparation protocols.  The non-adiabatic scheme switches suddenly from a standard 2D lattice to the HH lattice. The adiabatic protocol uses a quantum phase transition to the 3D Mott insulator as an intermediate step. For this, lattice beams are adiabatically ramped up in all three directions to 20 $E_r$ to enter the Mott insulating phase, after which the hyperfine state is flipped with a 0.29 ms RF sweep as done in the non-adiabatic scheme. The lattices are then ramped down to their final values while the Raman lattice is ramped up in 35 ms. Lifetimes of both methods are given in the top corner of each figure; the lifetime of the adiabatic approach is measured from the end of the 35 ms ramp. Errors are statistical and given by the exponential fit of the peak visibility.}
\end{figure}

Adiabatic methods usually require a pathway where the ground state is protected by a gap. For single particle states, this may involve matching the size of the unit cell between a topologically nontrivial lattice and a trivial lattice \cite{CooperPRA2014, BlochNAT2015}.  Instead, we explore the use of many-body gap of the bosonic Mott insulator state. Since the Bose-Hubbard model and the HH model with interactions have the same Mott insulating ground state for large U (assuming the high frequency limit $U\ll \Delta$), one can connect a trivial superfluid in the cubic lattice to the ground state of the HH model via two quantum phase transitions.  The first transition ``freezes'' out the original superfluid by entering the Mott insulator where the phases of the tunneling matrix elements can be changed to a non-zero flux configuration without adding entropy \cite{BlochSCI2013}. The second transition ``unfreezes'' the Mott state into the HH superfluid. Technically, this approach cannot be fully adiabatic due to the exact degeneracy of the ground states in the HH spectrum; a Kibble-Zurek model implies that this will lead to the spontaneous formation of domains \cite{ZollerPRL2005}. However, this will not create entropy beyond the randomization of domains.

We implement this scheme to load the ground state of the HH Hamiltonian. So far, we have found the adiabatic turn-on to be less robust against technical noise, leading to a higher shot-to-shot variability; however, the best images have identical visibility and a marginally better lifetime than that of the non-adiabatic state preparation method (Fig. 4). 

\begin{figure}	
\centering	
\includegraphics[width=\columnwidth]{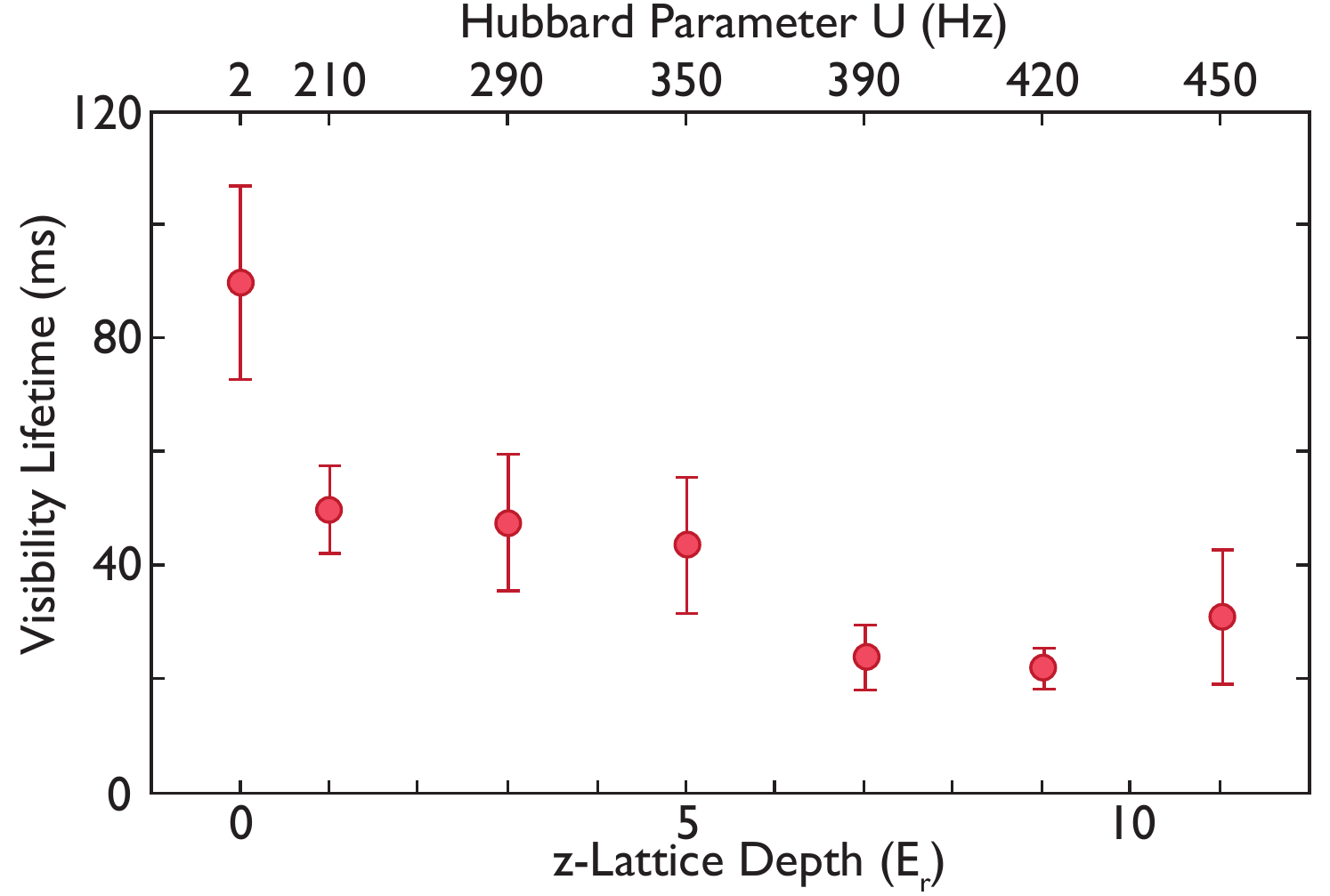}
\caption{HH model with strong interactions.  Shown is the lifetime of the visibility of the diffraction pattern versus z-lattice depth. The top axis shows the Hubbard interaction parameter U. All lifetimes are measured from the end of the 35 ms ramp exiting the Mott insulator. Uncertainty is given by the statistical error of the mean of five repetitions of the experiment, added in quadrature to uncertainty in the peak visibility fitting.}
\end{figure}

We can now use the lattice in the third dimension, required for the adiabatic loading protocol, to add stronger interactions to the physics of gauge fields. To do this, we keep the lattice depths and tunneling strengths along the x- and y-directions constant, but vary the depth of the z-lattice, thereby adjusting the Hubbard interaction parameter U from nearly zero to $\sim$450 Hz. With an estimated filling factor of $\sim$5, we can access interaction energies in excess of $h\times2\ \mathrm{kHz}$, which is close to the superfluid-to-Mott insulator transition in high magnetic fields \cite{OktelPRA2007}.

We realize HH superfluids with lattice depths from 0 to 20 $E_r$ and observe at least weak superfluid diffraction peaks (see Supplement). Coherence lifetimes for z-lattice depths up to 11 $E_r$ are presented in Fig. 5. The average visibility of the diffraction pattern is reduced at high z-lattices, but the lifetimes are reduced only by a factor of $\sim2-4$ from the lifetime of the weakly interacting superfluid at 0 $E_r$. However, we observe much larger shot-to-shot variation: some shots will have clearly visible diffraction peaks, while other attempts show no coherence at all. We attribute the variation to technical fluctuations, which seem to have an enhanced effect at higher 3D lattices.

In conclusion, we have observed Bose-Einstein condensation in a strong synthetic magnetic field. We have directly observed how the gauge potential breaks the symmetry of the original lattice. To reach the regime of strong interactions, we have demonstrated that the HH model can be implemented with reasonably long coherence times in the presence of both strong gauge fields and strong interactions. Therefore, this work lays the groundwork for further investigations of topologically nontrivial and/or highly frustrated states. Straightforward modifications of our setup such as the introduction of complex tunneling in the z-direction or pseudospins realize Weyl points \cite{DubcekARV2014}, the quantum spin Hall Hamiltonian, and ultracold atom topological insulators \cite{KetterlePRL2013b}. Recent technological advances in controlling and detecting ultracold atoms combined with this work offer a promising path toward realization of bosonic Laughlin states in optical lattices \cite{OktelPRA2007, LewensteinBOOK2012, SpielmanRPP2014, DalibardPRL2013}. The ability to realize strong gauge fields with strong interactions highlights the capability of ultracold atom systems to realize exotic states of matter with no known analogues in nature.

\section*{Acknowledgements}
We acknowledge Quantel Laser for the gift of an EYLSA laser for our main cooling and trapping light, Georgios Siviloglou and Yuri Lensky for experimental contributions, and Erich Mueller, Sayan Choudhury, Alan Jamison, and Misha Lukin, Sankar Das Sarma, Sid Parameswaran, Ehud Altman, and Eugene Demler for stimulating discussions. W.C.C. is grateful for the Samsung Scholarship. This work was supported by the NSF through grant PHY-0969731, through the Center for Ultracold Atoms, AFOSR MURI grant FA9550-14-1-0035 and ARO MURI grant No. W911NF-14-1-0003.

\section*{Methods}
The experiment begins with a nearly pure BEC confined in a crossed dipole trap in the $|1,-1\rangle$ hyperfine state with $\sim$$7 \times 10^{5}$ $^{87}$Rb atoms. We turn on a magnetic field gradient to levitate the atoms against gravity and simultaneously weaken the dipole traps by lowering the power to their final values. From here, the non-adiabatic and adiabatic sequences differ.

In the nonadiabatic sequence, the condensate is adiabatically loaded into a two-dimensional optical lattice composed of one vertical beam and one horizontal beam with lattice constants $532$ nm and depth of 11 recoil energies ($\text{E}_{\text{r}}$) in both the x- and y-directions, in the presence of a very weak Raman lattice ($<0.1$ $\text{E}_{\text{r}}$) with relative frequency detunings equal to the Bloch oscillation frequency $\text{3,420}$ Hz. The two beams are beat together on a beam cube located as close to the atoms as our optical setup allows, and the signal is used to stabilize the relative phase of the Raman beams.

Once a phase lock is achieved, we turn on a large tilt in the regime $J \ll \Delta \ll$ E$_{\text{gap}}$ by sweeping the frequency of an RF field in 0.29 ms to transfer all the atoms to the $|2,-2\rangle$ hyperfine state. Here $J$ is the bare tunneling energy in the lowest band, $\Delta = \text{3,420}$ Hz is the energy offset between adjacent sites, and E$_{\text{gap}}$ is the energy gap between the lowest band and the first excited band. Upon completion of the RF sweep, the initial system represents an uncoupled stack of 2D BEC's.

For topologically trivial superfluids, resonant tunneling is re-established with amplitude modulation of the vertical (x-axis) lattice with modulation frequency $\Delta$. For the 1/2 flux superfluid resonant tunneling is instead re-established by linearly ramping up both Raman beams in $0.58$ ms to $2\Omega/\Delta = 1.6$, where $\Omega$ is the two-photon Rabi frequency. After a variable hold time, all laser beams are switched off to allow 20 ms time-of-flight, followed by absorption imaging.

In the adiabatic sequence, the condensate enters the Mott insulating phase after lattices in all three directions are adiabatically ramped up to 20 $E_r$. One of the Raman beams is then ramped to its final value in 15 ms, after which the hyperfine state is flipped with a 0.29 ms RF sweep as explained previously. The lattices are then ramped down to their final values while the second Raman beam is ramped up in 35.09 ms, thereby adiabatically connecting the Mott insulator to the 1/2 flux superfluid. Again, time-of-flight absorption imaging is performed after a variable hold time.

\section*{Author Contributions}
All authors contributed to experimental work, data analysis, and manuscript preparation.

\section*{Additional Information}
The authors declare no competing financial interests.

%


\pagebreak
\onecolumngrid

\onecolumngrid
\section{Supplement}
\setcounter{figure}{0}

\subsection{1/2 Flux Superfluid Diffraction Pattern Fitting}

We quantify the visibility of the 1/2 flux superfluid diffraction pattern above the broad, thermal background by fitting one-dimensional line cuts of the column density in both the x- and y-lattice directions. To design a peak visibility fitter for the HH superfluid, one needs to account for multiple degrees of freedom associated with the new peaks that emerge, in addition to the trivial superfluid peaks. In addition, the uncontrolled phase of the micromotion leads to changes in the diffraction pattern, as seen in the paper. This is because the relative optical path length between the cubic lattice and the running wave Raman lattice at the position of the atoms is unstabilized. When designing a fitter, we must build in the flexibility in handling shot-to-shot changes in the peak locations on the CCD camera.

\begin{figure}[!ht]
\centering
\includegraphics[width=0.5\textwidth]{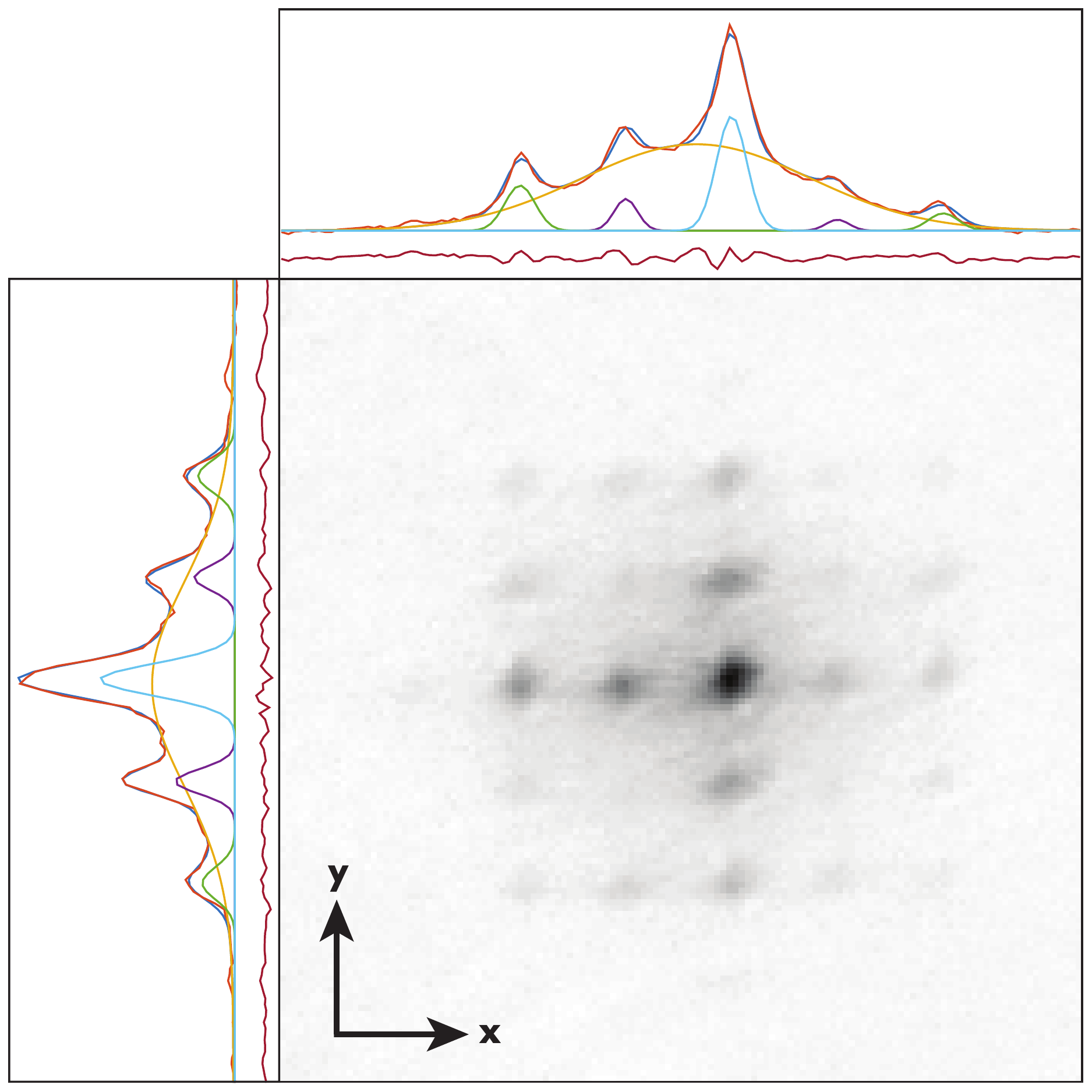}
\caption{An example fit for a 1/2 flux superfluid diffraction image after a hold time of 34 ms. The thermal fraction fit is shown in yellow with the components of the superfluid diffraction fit in cyan, purple, and green. Their sum is shown in dark blue laid over the raw data in orange. The residuals are shown in dark red. A visibility of $V = 0.40$ is derived from the amplitudes of this fit.}
\end{figure}

The fitter works by finding the center of mass of the image and taking the average of several rows of pixels in each direction around this center value. The total visibility is determined by least-squares fitting both the x and y line cuts to the sum of six gaussians -- one broad thermal background and five sharp diffraction peaks. The visibility of a given peak is defined as the amplitude of the coherent peak divided by the coherent amplitude plus the twice thermal background at the peak center. This is in analogy with Michelson's definition of optical fringe contrast. We define the final visibility as the average of visibility of the peaks in both the x- and y-directions.

Reliable fitting in such a high dimensional parameter space is robust when the fitting routine starts with accurate initial guesses. To generate accurate initial conditions, the image is first low pass filtered to remove spatial frequencies corresponding to the diffraction peaks and higher, and then fitted to a single gaussian function. The coefficients resulting from this fit become the initial conditions of the thermal fraction in the final fit. Next, the guess for the thermal component is subtracted from the raw lineout and all remaining local maxima are identified with a peak finding algorithm. The peaks are sorted by amplitude, and the highest five locations are chosen to represent the initial position and amplitude in the initial conditions for the superfluid diffraction peaks. With these input conditions, nonlinear least-squares fitting of the six gaussian peaks gave reproducible fits with small residuals. An example of a typical fit is shown in Figure 1.

\subsection{Bandmapping Procedure}

To quantify the population imbalance in the two minima of the HH spectrum, we developed a band mapping technique, in which laser beams are not switched off instantly but instead ramped down to zero smoothly before time-of-flight absorption imaging. The requirements for band mapping in the $\alpha = 1/2$ HH lattice are different from the standard bandmapping technique in two ways. First, the two Hofstadter sub-bands are connected at Dirac points and the total band gap is therefore zero, forbidding a globally adiabatic band mapping procedure.  Since the atoms are condensed, they are sufficiently localized to quasimomentum states that are far away from the Dirac points and are locally gapped. Second, there is a strong potential gradient present, leading to Bloch oscillations in the lab frame. For low lattice depths, the Bloch oscillations would cause Landau-Zener transitions at the zone edge. This effect is minimized by ramping down the lattice relatively quickly, in about one Bloch oscillation time. In the case of topologically trivial superfluids, all the lattices are linearly ramped down in $0.43$ ms. In the case of HH superfluids, Raman beams are linearly ramped down in $0.88$ ms first to avoid excitations associated with the Raman drive, and then the lattices are linearly ramped down in $0.43$ ms.

\subsection{Superfluid Diffraction Close to the Suerpfluid to Mott Insulator Transition}
\begin{figure}[!ht]
\centering
\includegraphics[width=0.6\textwidth]{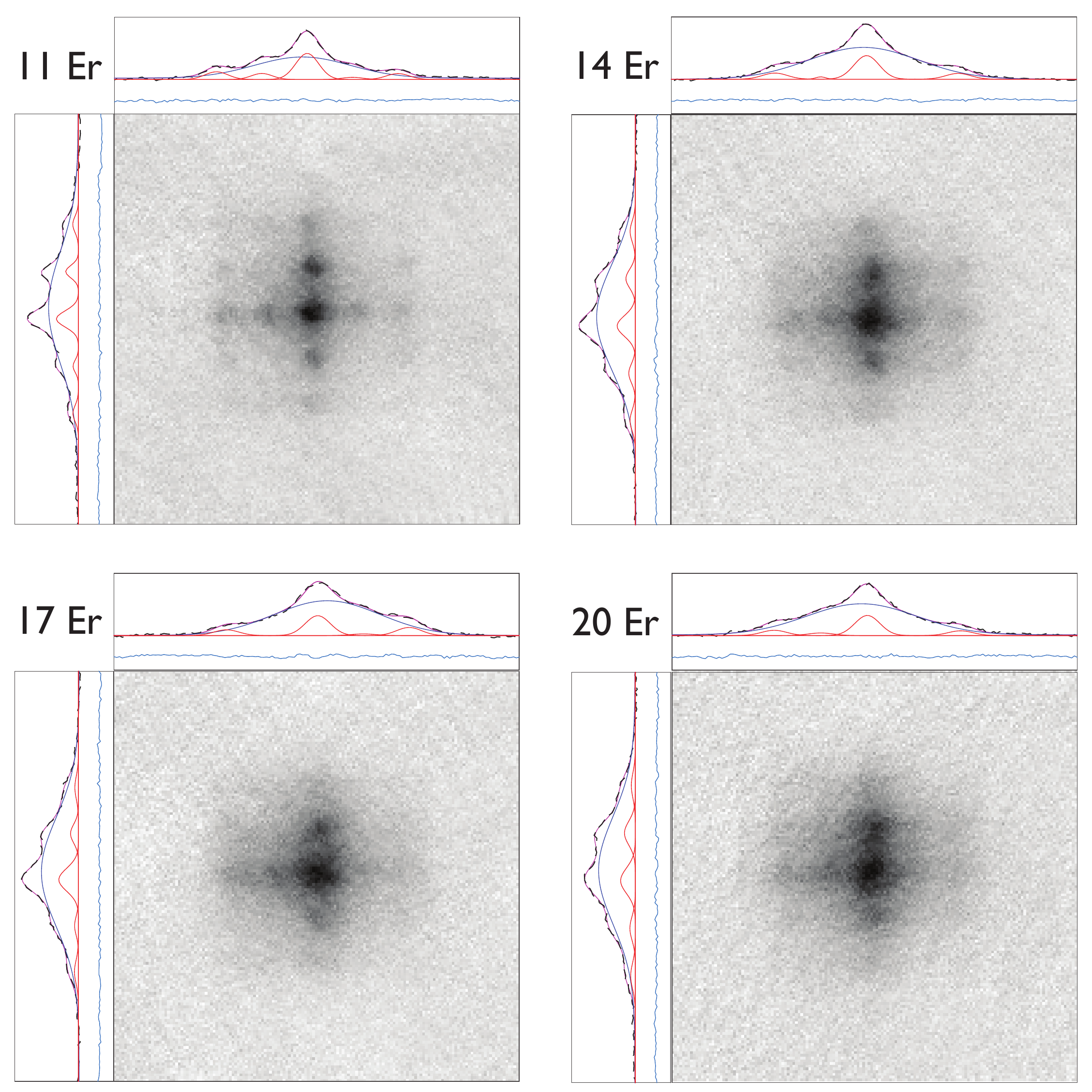}
\caption{1/2 flux superfluid diffraction pattern for increasing interaction strnength, controlled by the depth of the lattice in the z-direction.
}
\label{Image Gallery}
\end{figure}

As mentioned in the main text, we observe weak diffraction peaks in three-dimensional lattices for z-lattice depths up to 20 $E_r$ with x and y lattices held fixed at 11 $E_r$, demonstrating Bose-Einstein condensation in the HH lattice with strong interactions.  The images in Figure 2 represent a selection of shots immediately after the $\sim$35 ms ramp out of the Mott insulator state into the 1/2 flux superfluid. The doubling structure is clearly visible at lattice depths close to the Mott insulating transition for filling factors $n>2$.

No systematic studies were done for z-lattices greater than 11 $E_r$ since much larger shot-to-shot fluctuations were encountered, probably due to larger sensitivity to technical noise sources.

\subsection{Collisional instability of the ground state in tilted and driven lattices}

The observed decay of a Bose-Einstein condensate in tilted and driven lattices is much faster than the decay in standard lattices.  Although our lifetime seems to be limited largely by technical noise, we wish to discuss the limit given by the time dependent nature of the single-particle wave function representing the ground state of the Floquet Hamiltonian \cite{MuellerARV2014S,CooperPRA2014S}.  Choudhury {\em et al.} study the stability of a condensate in a tilted one-dimensional lattice with resonant amplitude modulation using a high frequency perturbation analysis.  The ground state of bosons in the corresponding Floquet Hamiltonian is a superfluid in a geometry with a lattice in one direction and free particle motion in the other two. The description of this system is simple yet nontrivial due to the metastability of the Wannier-Stark ladder and the time-modulated many-particle Floquet Hamiltonian. As pointed out in \cite{MuellerARV2014S}, one consequence of these two facts is that elastic collisions between atoms will deplete the condensate by using one or two quanta of the micromotion (equal to the offset energy $\Delta$ between neighboring sites) to create excitations.  This will cause decoherence and relaxation to higher temperatures.

There are two main processes contributing to this instability of the condensate. Wannier-Stark states -- which are mostly localized in a given well $n$ -- have small components in lower wells $n-1$, with amplitude $\propto 2J/\Delta$. This small component, which has energy $\Delta$ higher than the Wannier-Stark state localized in the well $n-1$, can collide with atoms in the lower lattice site, thus creating two excitations with energy $\Delta/2$ in the transverse direction (the creation of two exitations is necessary for conservation of transverse momentum) . The second mechanism involves the counter-rotating term of the phonon or photon assisted tunneling process. The co-rotating term enables resonant tunneling to a lower lattice site by stimulated emission of a phonon.  In the counter-rotating process, the atoms tunnels to a lower lattice site, but absorbs a lattice phonon.  Therefore, it has now an excess energy of $2\Delta$ which can create two transverse excitations where each of them has energy $\Delta$.

These two-body elastic collisions convert modulation or tilt energy into motion along the transverse directions.  The characteristic relaxation times of the two processes are given by $\tau_1$ and $\tau_2$:
$$
\frac{1}{\tau_1} \propto \frac{4J^2}{\Delta^2}\sum_k 2\pi\delta(E_k^{(0)}-\Delta/2)
$$
$$
\frac{1}{\tau_2} \propto \frac{K^2}{\Delta^2}\sum_k 2\pi\delta(E_k^{(0)}-\Delta).
$$
We can associate $\tau_1$ with a process that occurs even in the absence of modulation -- the rate is proportional to the tunnel coupling squared between neighboring sites, whereas the rate of the $\tau_2$ process is proportional to the square of the phonon induced tunneling rate.  Both decay processes are rather general and should apply to both tilted lattices and lattices modulated by superlattices.

As a test of these predictions, we implemented this Hamiltonian experimentally using the same experimental procedure described in the main text for resonant amplitude modulation. We measured the coherence and lower band population lifetimes as shown in Figure 3. Approximate values in our experiment were $gn = h \times 550-830\ \mathrm{Hz}, a_s = 5.03\ \mathrm{nm}, d = 254\ \mathrm{nm},\ K = h \times 10\ \mathrm{Hz}, J = h \times 30\ \mathrm{Hz},\ \mathrm{and}\ \Delta = h \times 3420\ \mathrm{Hz}$, with $gn$ as the chemical potential, $a_s$ the s-wave scattering length, $K$ laser-assisted tunnelling rate, $J$ the bare tunnelling rate, and $\Delta$ the energy tilt per lattice site. $d^{-1} = \int\ \mathrm{d}x\ \phi_x(x)^4$ is the size of the Wannier statein the x-direction $\phi_x(x)$. These give us the values of $1/\tau_2 = 0.003-0.005\ \mathrm{s}^{-1}$ and $1/\tau_1 = 0.13-0.20\ \mathrm{s}^{-1}$, which are longer than our observed lifetime.Uncertainty is driven by uncertainty in our density measurements due to redistribution during lattice ramp-up. However, during the long coherence lifetime, significant population loss was observed (due to inelatic two- or three-body collisions, or transfer to higher bands), and detailed further study is needed to identify the contributions of different processes to the decay rate.

\begin{figure}[!ht]
\centering
\includegraphics[width=0.6\textwidth]{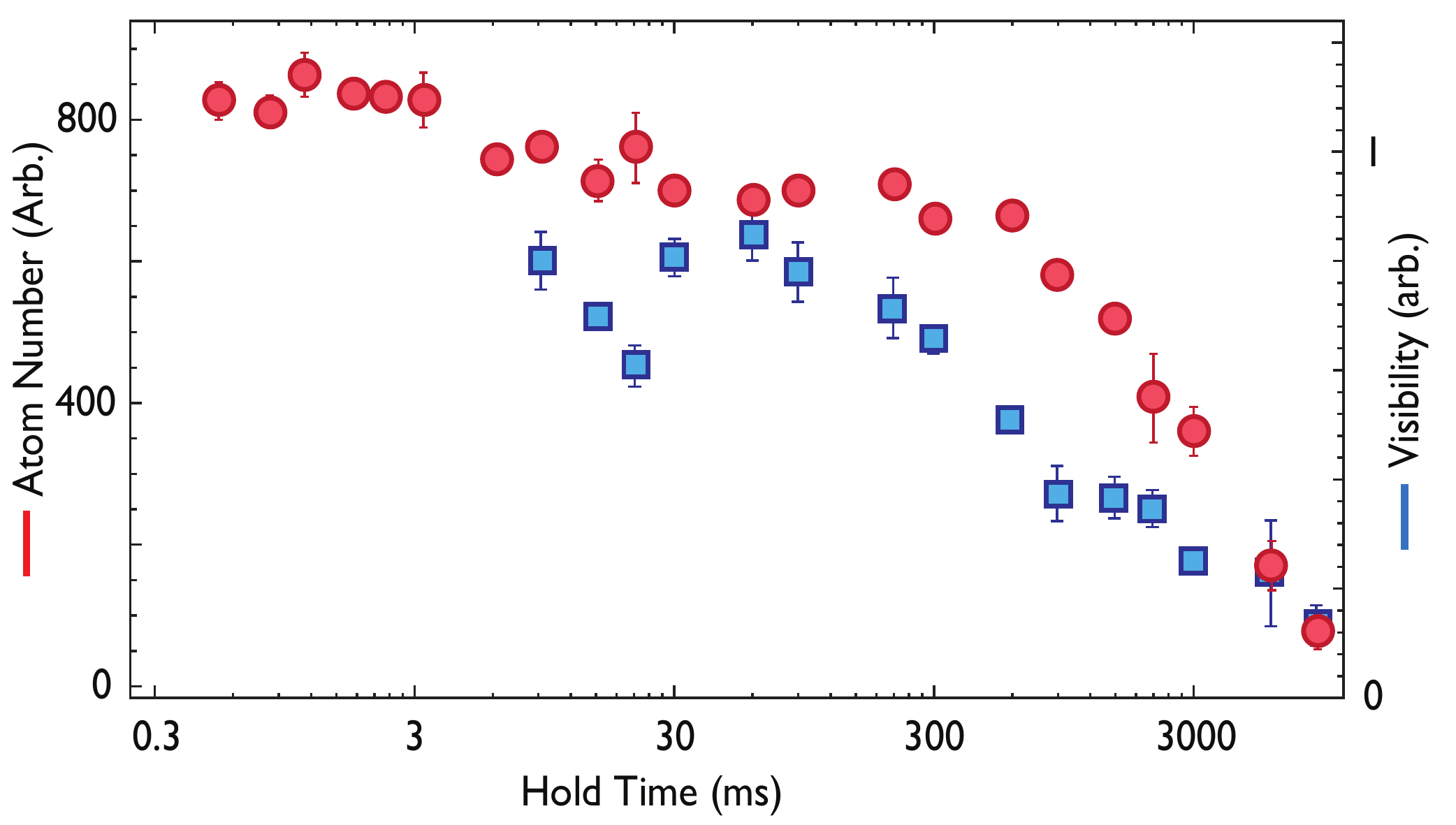}
\caption{Lifetime of a Bose-Einstein condensate in a 1D lattice with resonant amplitude modulation. Shown are the decay of the number of trapped atoms, and the decay of the visibility of the diffraction pattern observed after ballistic expansion.}
\label{1D AM}
\end{figure}

The mechanism proposed by Choudhury {\em et al.} appears to give a reasonable upper limit for collisional decay in 1D lattices.  We expect the same mechanism to explain at least part of the decoherence observed for both amplitude modulation in a two-dimensional lattice and the HH Hamiltonian. For a quantitative comparison, we modify the calculation in \cite{MuellerARV2014S} to reflect that a 2D lattice has only one orthogonal direction of free motion. For resonant modulation, in the Floquet basis, the Hamiltonian is
$$
H = \sum_k \epsilon_\mathbf{k}(t) b_\mathbf{k}^\dagger b_\mathbf{k} + \frac{g}{2V}\sum_{\mathbf{k_1,k_2,k_3}} b_\mathbf{k_1}^\dagger b_\mathbf{k_2}^\dagger b_\mathbf{k_3} b_\mathbf{k_4}
$$
with $\mathbf{k_4} = \mathbf{k_1}+\mathbf{k_2} -\mathbf{k_3}$, V as the system volume, and $b_\mathbf{k}$ and $ b_\mathbf{k}^\dagger$ as the annihiliation and creation operators for states with quasimomentum $\mathbf{k}$. The interaction parameter is given by
$$
g= \frac{4 \pi \hbar^2 a_s}{m} \lambda^2 \int\ \mathrm{d}x\ \mathrm{d}y \phi_x(x)^4 \phi_y(y)^4
 =\frac{4 \pi \hbar^2 a_s}{m} \frac{\lambda^2}{d^2}
$$
where $\phi_x(x)$ and $\phi_y(y)$ are the Wannier functions in the tilted (x) and untilted (y) lattice directions, $\lambda$ is the lattice spacing, and $d$ is the size of the Wannier state. The instantaneous single particle dispersion $\epsilon_\mathbf{k}(t)$ is given by
$$
\epsilon_k(t) = -2K \cos(k_x)-2J_y \cos(k_y) - 2K \cos(k_x-2\Delta t) - 2 J_x \cos(k_x - \Delta t) + \frac{\hbar^2 k_z^2}{2m}.
$$
where $J_x$ and $J_y$ are the bare tunnelling rates in the x and y directions, and $K$ is the effective tunnelling in the tilt direction. Following a similar argument, we find that the scattering rate is given by
$$
\frac{1}{\tau} = \frac{1}{\tau_2} + \frac{1}{\tau_1}
$$
$$
\frac{1}{\tau_2} = \frac{2 (gn/2)^2}{N\hbar}\frac{K^2}{\Delta^2}\sum_k 2\pi\delta(E_k^{(0)}-\Delta)
$$
$$
\frac{1}{\tau_1} = \frac{2 (gn/2)^2}{N\hbar}\frac{4J^2}{\Delta^2}\sum_k 2\pi\delta(E_k^{(0)}-\Delta/2)
$$
with the effective dispersion:
$$
E_k^{(0)} = 2K [1-\cos(k_x)]+2J_y[1-\cos(k_y)]+\sqrt{\frac{gn}{m} (\hbar k_z)^2+\left(\frac{(\hbar k_z)^2}{2 m}\right)^2}.
$$
Note that ref. \cite{MuellerARV2014S} uses a Hartree-Fock excitation energy with free particle dispersion. The high mean field energy in the one dimensional tubes created by a two dimensional lattice, $gn$, is similar in energy scale to the tilt frequency, and we thus use the full Bogoliubov spectrum for the energy of quasi-particles.

Since $K$ and $J_y$ are small, we neglect the dependence of $E_k^{(0)}$ on $k_x$ and $k_y$, giving
\begin{eqnarray}
\rho(\nu) & = & \sum_k 2\pi\delta(E_k^{(0)}-\nu) \\
\ & = &\frac{V}{\lambda^2}\int \frac{\mathrm{d}k_z}{2 \pi} 2\pi \delta\left( \sqrt{\frac{gn}{m} (\hbar k_z)^2+\left(\frac{(\hbar k_z)^2}{2 m}\right)^2}-\nu \right) \\
\ & = & \frac{Vm}{\hbar^2 \lambda^2} \frac{\nu}{\sqrt{(gn)^2+\nu^2}} \frac{\sqrt{\frac{2 \hbar^2}{m}}}{\sqrt{-(gn)+\sqrt{(gn)^2+\nu^2}}}
\end{eqnarray}
and thus
$$
\frac{1}{\tau_2} = \frac{gn}{\hbar} \frac{4 \pi a_s}{d} \frac{K^2}{\Delta^2} \frac{\Delta}{\sqrt{(gn)^2+\Delta^2}} \frac{\sqrt{\frac{\hbar^2}{2 m d^2}}}{\sqrt{-(gn)+\sqrt{(gn)^2+\Delta^2}}}
$$
$$
\frac{1}{\tau_1} = \frac{gn}{\hbar} \frac{4 \pi a_s}{d} \frac{4 J^2}{\Delta^2} \frac{\Delta/2}{\sqrt{(gn)^2+(\Delta/2)^2}} \frac{\sqrt{\frac{\hbar^2}{2 m d^2}}}{\sqrt{-(gn)+\sqrt{(gn)^2+(\Delta/2)^2}}}
$$

For our system, representative numbers are $gn = h\times 740-1740\ \mathrm{Hz}$, and $\Delta = h\times 3420\ \mathrm{Hz}$, $K = h \times 10\ \mathrm{Hz}$, $J = h\times 30\ \mathrm{Hz}$, $a_s = 5.03 \ \mathrm{nm}$, $d = 254 \ \mathrm{nm}$, and $m = 1.4\times10^{-25}\ \mathrm{kg}$, resulting in $1/\tau_2 = 0.006-0.014\ \mathrm{s}^{-1}$ and $1/\tau_1 = 0.29-0.67\ \mathrm{s}^{-1}$. This is much smaller than the decay rates measured in the experiment.

\subsection{Loss Processes in the 3D HH Model}

\begin{figure}[!ht]
\centering
\includegraphics[width=0.48\textwidth]{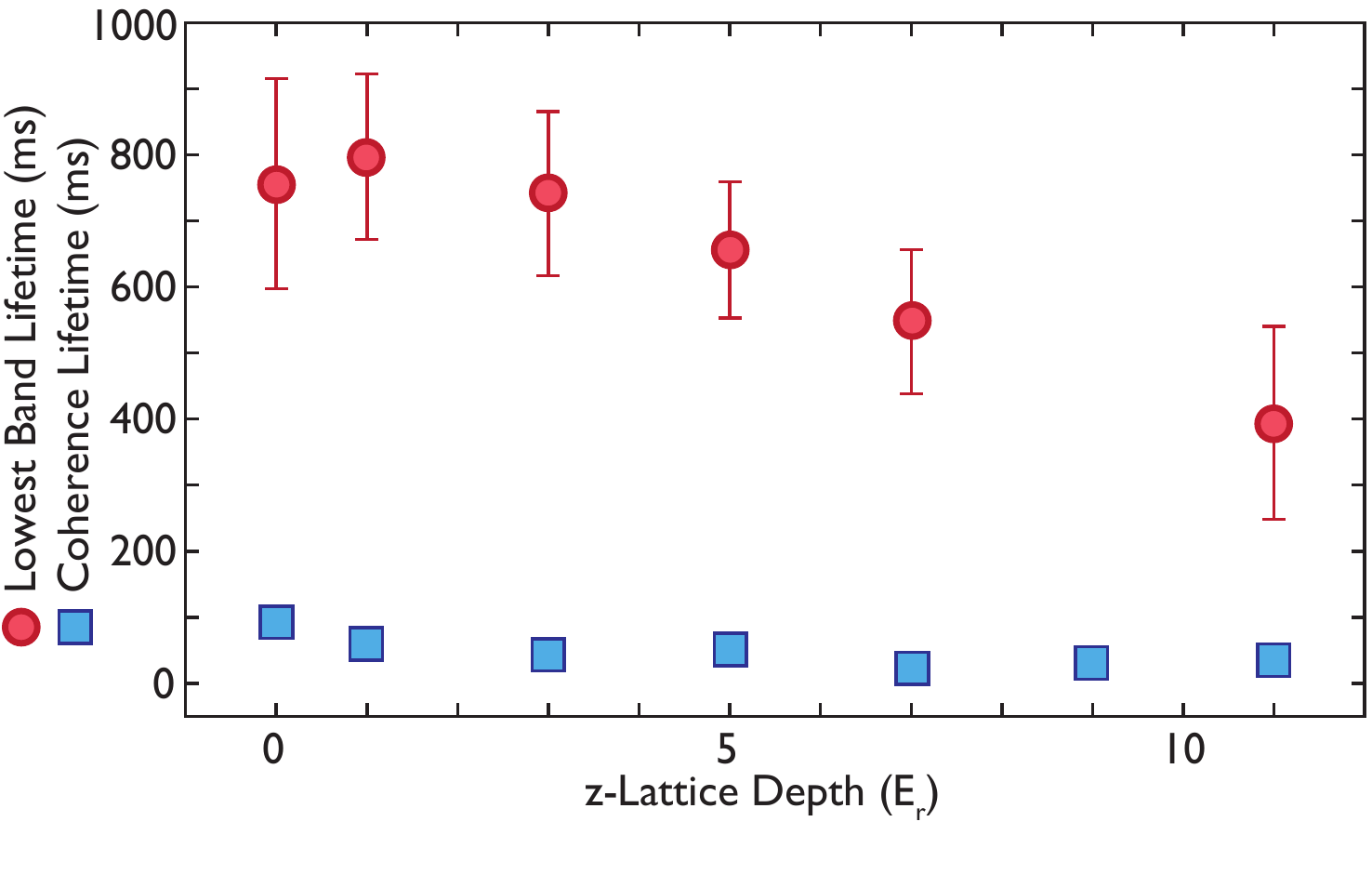}
\caption{Lifetime of the total atom number (red) in the lowest band of the 3D lattice as a function of the z-lattice depth. For comparison, the coherence lifetime (blue) as a function of lattice depth is shown on the same plot. In the main section of the paper the coherence data is displayed as a function of the calculated Hubbard interaction parameter, U.}
\end{figure}

By monitoring the loss of atoms we can show that certain processes are negligible for the loss rate of the superfluid state.  Figure 4 shows the lifetime of the population in the lowest band of our lattice -- or equivalently the total atom number lifetime, since the strong tilt will quickly remove population promoted to higher bands due to Landau-Zener tunneling to the continuum.  The lifetime for atom loss from the lower band is about an order-of-magnitude longer than the coherence lifetime and shows a slight decrease with increasing lattice depth.  For comparison, a BEC in an untilted lattice exhibits coherence times exceeding one second in deep lattices.

We conclude that the population loss seen in Figure 4 is dominated by some process beyond heating in standard lattices (spinflip collisions,  magnetic field noise, lattice intensity noise, and beam pointing noise).  However, the dominant decay process for the BEC in the HH lattice are excitations within the lowest band.

A 3D lattice should be helpful for reducing collisional heating. Qualitatively, in a Fermi's golden rule picture of heating, the decay rate depends on the dispersion relation of the transverse modes \cite{MuellerARV2014S}.  Thus, if a 3D lattice has a bandgap at the energies $\Delta$ and $\Delta/2$, at which quasiparticles can be created, we would expect to see a ``gapping out'' of collisional heating and a corresponding increase in the lifetime upper limit. Since we do not see a qualitative change in lifetime when changing the bandgap, we conclude that, similarly to a 2D lattice, technical fluctuations are the likely limit of the lifetime.

\end{document}